# REFERENCE-LESS DETECTION, ASTROMETRY, AND PHOTOMETRY OF FAINT COMPANIONS WITH ADAPTIVE OPTICS

Szymon Gladysz [1]

Department of Experimental Physics, National University of Ireland, Galway, University Road, Galway, Ireland

Julian C. Christou

Gemini Observatory, 670 North A'ohoku Place, Hilo, HI 96720, USA

ABSTRACT

We propose a complete framework for the detection, astrometry, and photometry of faint companions from a sequence of adaptive optics (AO) corrected short exposures. The algorithms exploit the difference in statistics between the on-axis and off-axis intensity of the AO point-spread function (PSF) to differentiate real sources from speckles. We validate the new approach and illustrate its performance using moderate Strehl ratio data obtained with the natural guide star AO system on the Lick Observatory's 3 m *Shane Telescope*. We obtain almost a 2 mag gain in achievable contrast by using our detection method compared to $5\sigma$ detectability in long exposures. We also present a first guide to expected accuracy of differential photometry and astrometry with the new techniques. Our approach performs better than PSF-fitting in general and especially so for close companions, which are located within the uncompensated seeing (speckle) halo. All three proposed algorithms are self-calibrating, i.e., they do not require observation of a calibration star. One of the advantages of this approach is improved observing efficiency.

## 1. INTRODUCTION

After the discovery of more than 300 exoplanets by indirect methods such as radial velocity or astrometry, direct imaging of these objects is the obvious next step. The next generation of extremely large telescopes and extreme adaptive optics (ExAO) dedicated to this specific science task are currently entering the phase of detailed planning and design (Kasper et al. 2008). The two ExAO systems, which will come online at the beginning of the next decade, are the Gemini Planet Imager (Macintosh et al. 2008) to be installed at the Gemini South telescope, and the SPHERE instrument which will be located at the focus of the Very Large Telescope (Beuzit et al. 2006). Both instruments will combine high-order AO correction, novel coronagraphic designs, and noise suppression modules to provide unprecedented contrast levels, e.g., $10^{-8}$ at a separation of $0\rlap{.}''5$ in 1 hr exposure (Macintosh et al. 2008).

Even though detection limits for current high-contrast AO systems are usually quoted together with an exposure time, empirical studies have shown that contrast does not improve with integration times longer than about 1 minute (Marois et al. 2005; Masciadri et al. 2005; Hinkley et al. 2007). This is because contrast is limited by static or quasi-static speckle noise which is not sensed, and therefore not corrected for, by the standard AO techniques. This particular noise contribution often arises after the wavefront sensor (noncommon-path errors), so static speckles stand out against the AO-corrected halo and masquerade as faint sources. Recent papers by Itoh et al. (2006) and Janson et al. (2006) have shown how difficult it is to obtain reliable detections in the presence of static speckles.

---

[1] Current address: European Southern Observatory, Karl-Schwarzschild-Straße 2, 85748 Garching, Germany.

As mentioned before, both planned high-contrast instruments will incorporate speckle suppression modules. Proposed techniques are based on the concept of point-spread function (PSF) subtraction and utilize PSF estimates provided by the on-sky rotation (Marois et al. 2006), as well as spectral (Racine et al. 1999; Marois et al. 2005; Sparks & Ford 2002) and polarization-based (Gisler et al. 2004) discrimination between the light coming from the parent star and the companion. The first two methods are already being used in surveys dedicated to finding faint companions (Lafrenière et al. 2007; Biller et al. 2006). The images they yield still contain static speckles (at a lower brightness level than in the direct images) due to errors in PSF estimation. These errors arise due to the inherent sensitivity of PSF subtraction to changes in seeing, mechanical flexures and PSF drifts, or the introduction of extra imaging channels (Biller et al. 2004; Marois et al. 2005; Cavarroc et al. 2006). Another alternative is to use a focal plane wavefront sensor (Guyon 2006) that senses *all* the aberrations downstream from the coronagraph and the first-stage AO system, or interferometric real-time calibration of the time-averaged wavefront, which will be an integral part of GPI (Macintosh et al. 2008).

We believe that the capability of current AO systems can be greatly enhanced by exploiting the statistical information present in multiple images of the same object. Just as the spectral signature of a planet is used to isolate its signal from the starlight (Marois et al. 2005), so can statistical information about its intensity – which is simply the scaled and shifted copy of the on-axis intensity in an isoplanatic approximation – be utilized for the same task. The advantage of the "stochastic speckle discrimination" methods, first proposed by us (Gladysz & Christou 2008), is that they are sensitive to *any* objects located close to bright stars, because these techniques only depend on the properties of the image formation process. Their application is therefore not limited to cases of exoplanets with specific atmospheric composition; neither is it restricted by lower limit on the angular separation (Thatte et al. 2007). We should mention the work by Labeyrie (1995) here. His "dark speckle" method was the first attempt to use intensity statistics to reveal exoplanets, but the technique suffers from the same static-speckle problem mentioned previously. "Dark speckle" requires intensity modulation, while the method proposed here does not. Recently, a technique for such modulation carried out in the pupil plane has been proposed (Ribak & Gladysz 2008), solving some of the problems pertaining to "dark speckle."

In Section 2, we describe statistical distributions of on-axis and off-axis (speckle) intensity. We also discuss our previous work on "stochastic speckle discrimination" and propose a new algorithm for detection. In Section 3, we test the algorithm's performance on real and simulated images of binary stars obtained with the Lick Observatory natural guide star AO system on the 3 m Shane Telescope. We arrive at the dynamic range plot for the proposed method. Then Section 4 presents a simple extension of the detection algorithm to the problem of differential astrometry in the presence of speckle noise. We formulate the concept of one-dimensional "distribution deconvolution" and show how it can be used to arrive at accurate differential photometry in Section 5. Concluding remarks follow in Section 6.

2. INTENSITY STATISTICS IN AO-CORRECTED IMAGES

This section is only a brief introduction to the subject of speckle statistics. For a complete review the reader should consult Cagigal & Canales (1999), Soummer et al. (2007), and Gladysz et al. (2008a) for the particular case of the image central point.

The speckle intensity distribution or – using formal nomenclature – probability density function (PDF), for any location (except the center, as will be explained later) in the image plane is described by the modified Rician function

$$p(I) = \frac{1}{I_s} \exp\left(-\frac{I + I_c}{I_s}\right) I_0 \left(\frac{2\sqrt{I}\sqrt{I_c}}{I_s}\right), \qquad (1)$$

where $I_c$ corresponds to the intensity produced by the deterministic (constant) part of the wavefront, and $I_s$ corresponds to the halo produced by random intensity variations. $I_0$ is the zero-order modified Bessel function of the first kind. It should be emphasized that the parameter $I_c$ relates to the static wavefront component and not only the perfect (flat) wavefront. It therefore captures *all* static aberrations present in the optical system. The parameters $I_c$ and $I_s$ are related to the expected value $E(I)$ and variance $\sigma^2_I$ of intensity through the following equations:

$$E(I) = I_c + I_s, \qquad \sigma_I^2 = I_s^2 + 2I_c I_s. \qquad (2)$$

Additional intensity variance associated with the Poisson statistics can be added to Equation (2) in the photon-counting regime. The expression for $\sigma^2_I$ then becomes

$$\sigma_I^2 = I_s^2 + 2I_c I_s + I_c + I_s. \qquad (3)$$

In this paper, we will not deal with such situations and therefore the form of Equation (2) is sufficient for our purposes.

The PDF given by Equation (1) reduces to the exponential distribution when no real-time correction is provided, i.e., $I_c \ll I_s$. When the wavefronts become more coherent due to the action of AO, the PDF becomes more symmetrical, i.e., the skewness shifts from highly positive values (high-end tail) toward zero. Figure 1 shows the modified Rician PDF for three levels of the static component $I_c$, i.e., for three levels of wavefront coherence.

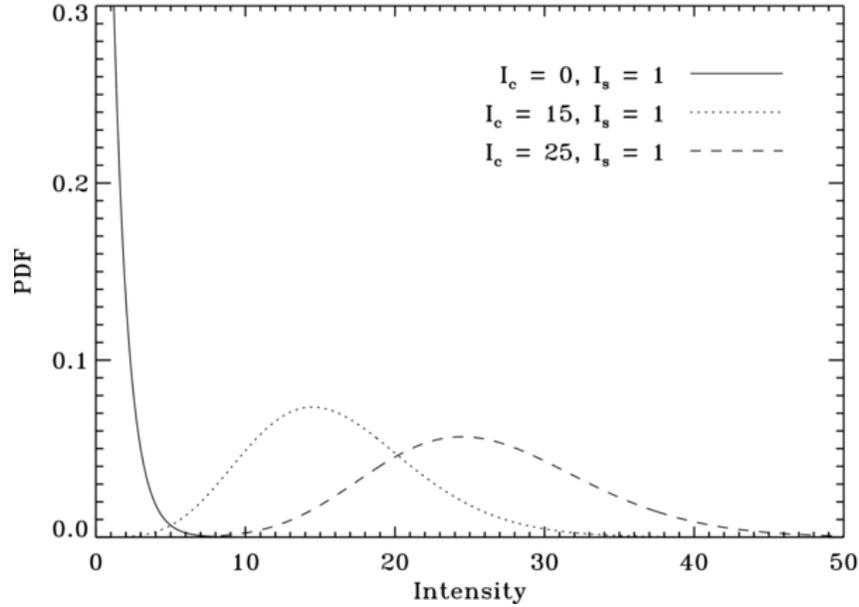

Figure 1. PDF of speckle intensity given by Equation (1) for three different deterministic intensity levels $I_c$. The ordinate axis was shortened for better visibility of all three distributions.

We found that Equation (1) is not valid at the central point of the focal plane. The normalized, i.e., divided by the peak value of the diffraction-limited image, on-axis intensity – the Strehl ratio (SR) – was shown to follow this PDF (Gladysz et al. 2008a):

$$p_{SR}(\text{sr}; k, \theta, \mu) = \frac{p_{\sigma_\phi^2}(-\ln \text{sr}; k, \theta, \mu)}{\text{sr}}, \quad (4)$$

where SR denotes the random variable with possible values sr, and $p_{\sigma_\phi^2}(\ )$ is the distribution of the phase variance described by the gamma model:

$$p(\sigma_\phi^2; k, \theta, \mu) = \frac{\left(\frac{\sigma_\phi^2 - \mu}{\theta}\right)^{k-1} \exp\left(-\frac{\sigma_\phi^2 - \mu}{\theta}\right)}{\Gamma(k)\theta} \quad \text{for } \sigma_\phi^2 \geq \mu, \quad (5)$$

where $k > 0$ is the shape parameter, $\theta > 0$ is the scale parameter, $\mu$ is the location parameter, and $\Gamma(x)$ denotes the gamma function. The value of $k$ is related to the number of independent phase patches in the wavefront, $\theta$ is related to the mean phase variance, and $\mu$ corresponds to the static aberrations in the wavefront (shifting the phase variance PDF toward higher values; Gladysz et al. 2008a).

In Figure 2, the distribution of the Strehl ratio is plotted for three levels of turbulence strength, as quantified by the Fried's parameter, $r_0$. We model the Lick AO system: telescope's diameter = 3 m, 61 actuators in total, servo bandwidth = 50 Hz, and we assume a relatively bright star $m_V = 9$ so that the AO frame rate is 500 Hz. The parameters of the AO system and the atmosphere can be related to the mean phase variance via the error budget equations (Gladysz et al. 2006). In our calculations, we included the fitting error, the bandwidth error, the time delay error, and the measurement error. The calculations overestimate the Strehl ratio because we did not include other sources of error, for example, static aberrations in the optical system. It should be noted that Equation (4) is only valid in the moderate and high Strehl ratio regime ($0.3 < \text{SR} < 1$).

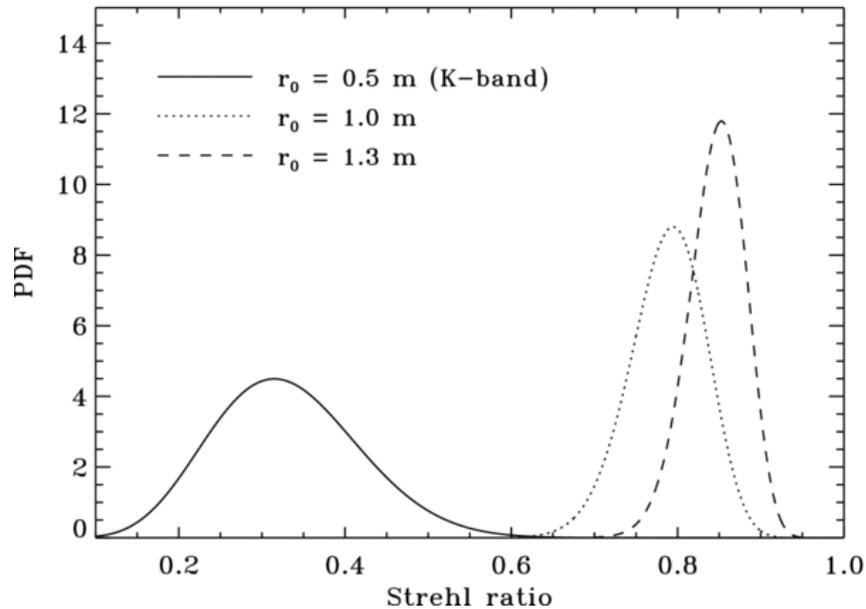

Figure 2. Theoretical distributions of the Strehl ratio for three levels of turbulence strength. An AO system similar to Lick was modeled.

Equations (4) and (5) pertain to the distribution of the Strehl ratio, but in order to make comparisons with the speckle PDF – Equation (1) – we need to derive an expression for the distribution of "raw" on-axis intensity. Form of this on-axis PDF will also be important in Section 5, where we develop a new photometric technique. Strehl ratio is the measured on-axis intensity divided by the diffraction-limited peak intensity when observing a point source. Just like in our previous work (Gladysz et al. 2008a) we define the "instantaneous" SR as the ratio of the peak value of a single short exposure $I$ divided by the peak value of the diffraction-limited image $I^*$:

$$\mathrm{SR} = \frac{I_{\mathrm{peak}}}{I^*_{\mathrm{peak}}}, \qquad (6)$$

where we assumed the perfect image $I^*$ was initially normalized to have the total power matching that observed for a given star. This is best done by normalizing the image $I^*$ to have the same power as the averaged shift-and-add (SAA), or long exposure image (Gladysz et al. 2006). The average images have a better signal-to-noise ratio (S/N) and therefore estimates of the total flux are generally more accurate than for short exposures, although this flux normalization can be carried out only when the conditions remain photometric during the recording of the entire data cube.

A distinction is sometimes made between on-axis and peak intensity (in the presence of image motion) but in our work we assume perfect re-centering of the frames, so that the pixel containing peak intensity is the central pixel. Given these two conditions it can be stated that on-axis intensity for each short exposure in a data set is related to that frame's SR by a single number which is constant for the entire data set, namely, the peak intensity of the flux-normalized perfect PSF. With this in mind one can find the distribution of on-axis intensity by transforming the SR PDF:

$$p(I) = \frac{p_{\mathrm{SR}}(I/I^*_{\mathrm{peak}})}{I^*_{\mathrm{peak}}}. \qquad (7)$$

The final expression for the on-axis intensity PDF contains four parameters

$$p(I; k, \theta, \mu, I^*_{\mathrm{peak}}) = \frac{\left(\frac{-\ln(I/I^*_{\mathrm{peak}}) - \mu}{\theta}\right)^{k-1} \exp\left(\frac{\ln(I/I^*_{\mathrm{peak}}) + \mu}{\theta}\right)}{\Gamma(k)\theta I}. \qquad (8)$$

It is intuitively obvious that the multiplication of a random variable by a scalar does not change the shape of its distribution and this is precisely what Equation (8) shows. This distribution of AO on-axis intensity has the same shape as the SR PDF and this shape is significantly different from that of speckle PDF in the moderate to high SR range, where skewness of the on-axis distribution is negative. We propose to use this morphological difference between distributions to discriminate between faint companions and static speckles.

2.1. Stochastic Speckle Discrimination

We studied the difference between on-axis and off-axis PDFs in our previous work (Gladysz & Christou 2008) by directly comparing estimates of the distributions. The Anderson-Darling two-sample test was used to quantify the deviation from speckle statistics. The proposed approach to detection assumed that the $p$-value resulting from the Anderson-Darling test could be used to enhance the confidence that a given objectlike feature in an image really is a companion and not just a static speckle. It was found that in order to obtain a reliable estimate of the probability of false-alarm (PFA) the test has to be executed for every location within the speckle "cloud" making this approach very computationally intensive.

Stochastic speckle discrimination (SSD) outlined briefly above required the computation of the parameters $I_c$ and $I_s$ for all the time series in a peak-shifted data cube (pixels in the focal plane), so that subsequently only time series with similar $I_c$ and $I_s$ were compared. This computation was done using the method of moments, whereby the unobservable population moments are equated to their sample estimators. Equation (2) is then inverted to produce estimates of $I_c$ and $I_s$:

$$I_s = E(I) - \left(E(I)^2 - \sigma_I^2\right)^{0.5}$$
$$I_c = E(I) - I_s. \qquad (9)$$

We observed that for the pixels where the artificial companions were located, the estimation of the parameter $I_c$ yielded values higher than for the speckles of similar magnitude. Conversely, the value of $I_s$ was significantly lower for the companion's peak pixel than for the speckles. This is easy to explain heuristically. On-axis intensity is the squared Fourier coefficient corresponding to the DC wavefront component, or the coherency of the wavefront. The goal of AO is to stabilize the wavefront (and Strehl ratio). $I_c$, which captures the amount of coherence in a wavefront, will be large at the center of the PSF. On the other hand, the parameter $I_s$ is a measure of the local intensity variations. For two pixels in the focal plane with similar mean intensity, one containing the companion and the other containing speckle, the intensity variations will be greater for the latter.

With this in mind we can design image post-processing techniques which transform spatial intensity variations (SAA images) into maps of local statistics. We tested a number of transformations, and in this paper we report on the two which gave best results:

$$I_{\text{SAA}}(x, y) \mapsto \frac{I_s(x, y)}{I_c(x, y)} \qquad (10)$$

$$I_{\text{SAA}}(x, y) \mapsto \frac{I_s(x, y)}{I_c(x, y)} m_3(x, y), \qquad (11)$$

where $m_3$ is the skewness sample estimator, defined for each location $(x, y)$ as

$$m_3 = \frac{1}{N} \sum_{j=0}^{N-1} \left(\frac{I_j - E(I)}{\sigma_I}\right)^3 \qquad (12)$$

and $N$ is the number of frames in a data cube. The two transformations given by Equations (10) and (11) lead to two new representations of an object where centers of the PSFs have low values compared to the speckles which now have relatively high values. This difference is the basis of our new speckle discrimination method. The latter transformation is dictated by the fact that the on-axis intensity distribution given by Equation (8) possesses negative skewness. Addition of intensities from two independent random processes, i.e., speckle and peak of the companion's PSF, leads to a convolution of the corresponding PDFs. Therefore, the negative skewness of Equation (8) will gradually shift toward zero, and then will become positive, when the relative magnitude of the speckle contribution increases, i.e., the observer searches for fainter and fainter companions. One should not expect a negatively skewed intensity histogram at the location of the suspected source. Still, the skewness should be smaller than in the case of pure speckles with similar mean intensity. For real sources multiplication by skewness should decrease the ratio of $I_s$ and $I_c$ even further. As will be shown in the next section, in reality this extra factor is not helpful because of the noisy estimation of the skewness.

3. STATISTICS-BASED DETECTION OF FAINT COMPANIONS

In the sections that follow we use exclusively short-exposure, moderate-Strehl ratio data from our observing campaign at Lick Observatory. Over three years we observed 50 single stars and seven double stars. The integration times of the IR Camera for Adaptive Optics at Lick (IRCAL) used in the observations were either 22 ms or 57 ms and we collected ~$10^4$ exposures per target. All images were taken in the *K* band, where the diffraction limit of the 3 m telescope is Nyquist-sampled by the detector pixels ($\lambda/D = 151$ mas). With Nyquist-sampled data it was possible to register individual short exposures via iterative Fourier shifting to produce SAA images. The short-exposure mode of the IRCAL camera (Fitzgerald & Graham 2006) adds an estimated $30 e^-$ rms readout noise, compared to only $12 e^-$ rms in the standard mode. The measured Strehl ratios in these SAA images were in the range 0.3 - 0.55. For the details of the observations and data reduction see Gladysz et al. (2006).

3.1. Simulated Faint Companions

In order to test the new approach we generated observations of artificial companions by scaling and shifting a single-star data set. This process scales the mean value and standard deviation of each pixel's time series by the same number. The problem with this approach arises in the photon-counting regime. For example, a sequence with the mean value equal to 100 counts and the standard deviation of 10 counts, divided by 10, gives a time series with the mean of 10 and the standard deviation of 1. But from the Poissonian statistics alone, for this new signal standard deviation should be 3.16 counts. This argument shows the danger of simulating faint-companion observations by scaling and shifting single-star frames. Therefore, we needed to check the validity of this approach. We measured standard deviations of on-axis intensity for all data sets and plotted them versus mean values. A clear linear relationship with a slope of 0.2 was identified down to 20 counts (lowest observed on-axis flux). In our observations the gain was set to 10 so this low-end limit corresponds to $200 e^-$. For the tests described below we selected the highest flux data set (star HD 216756, $m_K = 4.8$, exposure time = 57 ms, filter FWHM = 0.32 μm, SR = 0.51). This selection was dictated by the intention to simulate systems with large magnitude difference down to $200 e^-$, where we get valid results because of the measured linear dependence of intensity variations. This linear relationship justifies scaling and shifting down to $\Delta m$ of 7.5 in the case of HD 216756. It also confirms that we are not in photon nor read noise limited regime over the dynamic range explored, as we have posited previously (Section 2).

Even though HD 216756 is cataloged as a single star we observed a companion, or background object, $1''\!.4$ away from the star. This does not influence the presented results. In fact, this is quite fortunate as it allows the new approach to be tested for the detection of real and simulated sources simultaneously. Figure 3 show the SAA image of HD 216756 in false color and square-root intensity scale. Note the static speckles – some of them persisted for months.

A simulated faint companion ($\Delta m = 4.25$, brightness ratio = 50) was added to the above image 5 pixel down and 3 pixel left of the center (separation = $0''\!.45$, just below the Airy ring; the central panel in Figure 3). At this location the data cube had low intensity, so the obtained composite PDF (of the random variable equal to the sum of the speckle and the companion intensities) was morphologically similar to the central location's PDF. A 1 pixel wide annulus of radius equal to the star-companion separation was used to compute the standard deviation of speckle noise. $3\sigma$ noise metric divided by the star's peak intensity was equal to 0.015. The simulated companion-to-star brightness ratio (0.02), while higher than $3\sigma$, does not result in reliable visual detection in the direct image. This example shows why some authors have suggested that $3\sigma$ is not an appropriate criterion for describing the detectability of faint companions in speckle-dominated AO observations (Kasper et al. 2007; Marois et al. 2008). Figure 4 shows the SAA image together with the maps corresponding to the two transformations given by Equations (10) and (11). Locations dominated by Gaussian noise, where the computation of $I_s$ and $I_c$ with Equation (9) proved impossible because $\sigma^2_I > E(I)^2$, were replaced with a constant (white pixels outside the speckle cloud in Figure 4).

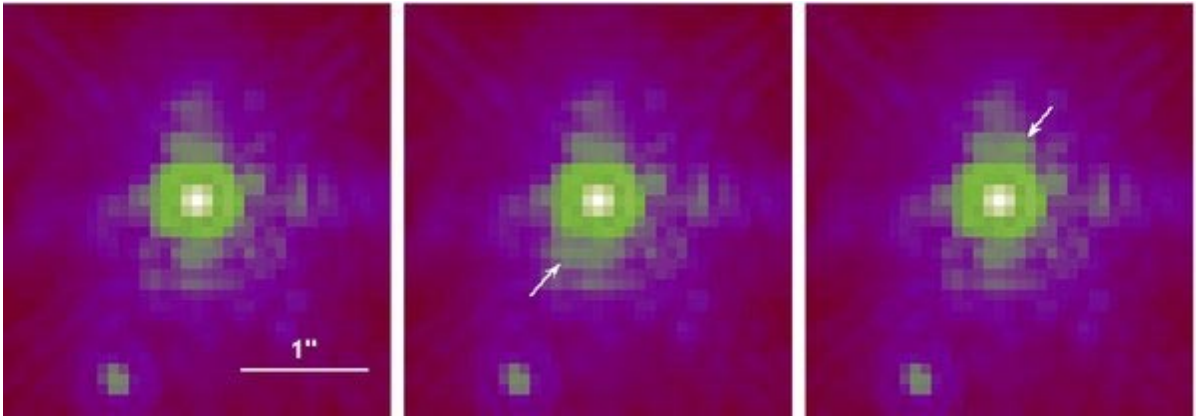

Figure 3. Left: SAA image of the star HD 216756 in square-root intensity scale and false color. Center: simulated companion ($\Delta m = 4.25$) was positioned on the zero of the speckle distribution. Right: same companion located on top of a static speckle.

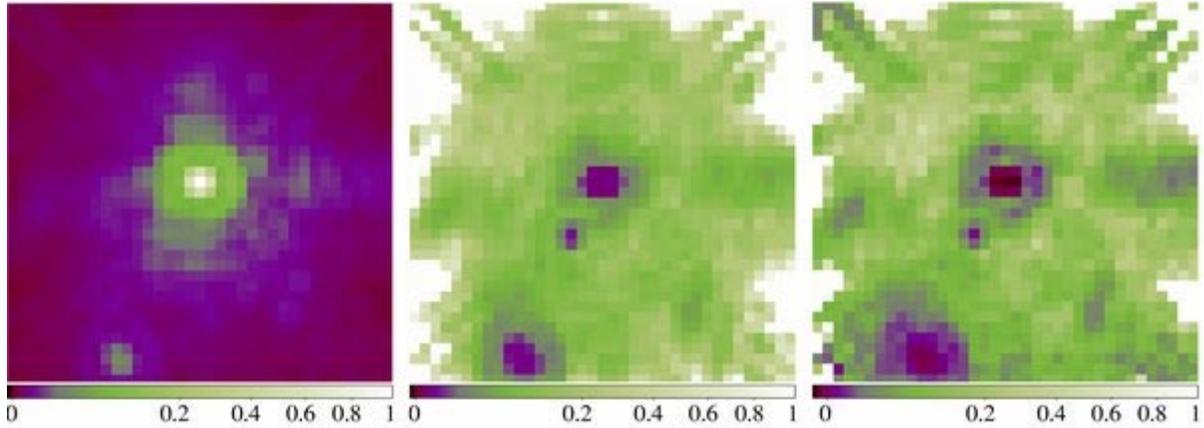

Figure 4. Simulated triple-star system with the faint companion ($\Delta m = 4.25$, separation = $0\rlap{.}''45$) located on the zero of the speckle distribution – 5 pixel down and 3 pixel left of the center. SAA image of the system (left), $I_s/I_c$ (center), $m_3 \cdot I_s/I_c$ (right). The intensity values are displayed in the square-root scale. Color scale of the first panel is expressed in fraction of the peak pixel intensity. In the central and right panels, the color scale is given in absolute units.

While only the two bright stars can be visually identified in the SAA image, the two transformations reveal the second companion, which can now be easily seen against the halo of speckle statistics. What is remarkable is that these transformations – especially Equation (10) as shown by the central panel in Figure 4 – yield very smooth maps compared to the direct images. The values of $I_s/I_c$ and $m_3 \cdot I_s/I_c$ seem to be almost constant across the focal plane. This gets rid of the static speckle problem. In our previous paper (Gladysz & Christou 2008), we postulated that the accurate computation of the PFA is made very difficult, if not impossible, by the presence of static speckles. If the PSF is highly anisotropic the ergodic hypothesis cannot be used as a justification for using spatial moments of intensity to calculate the PFA. The new statistical representation of an object that we propose here can be used to compute S/N and PFA spatially, which makes it an excellent tool for detection. As observed by Soummer et al. (2007) the modified Rician PDF breaks down not only for the central pixel but also for its local neighbourhood. This is clearly visible in the two statistical maps, where the central nine pixels are deep purple.

Next, we positioned the same faint companion (Δ*m* = 4.25) on top of a static speckle (five pixels up, two pixels right of the center; rightmost panel in Figure 3). The separation was slightly smaller (~0″.41), and the statistics were affected more by speckle than in the previous case, therefore the visual results were worse, as seen in Figure 5. In particular the contrast in the right panel in Figure 5 is very low compared to Figure 4.

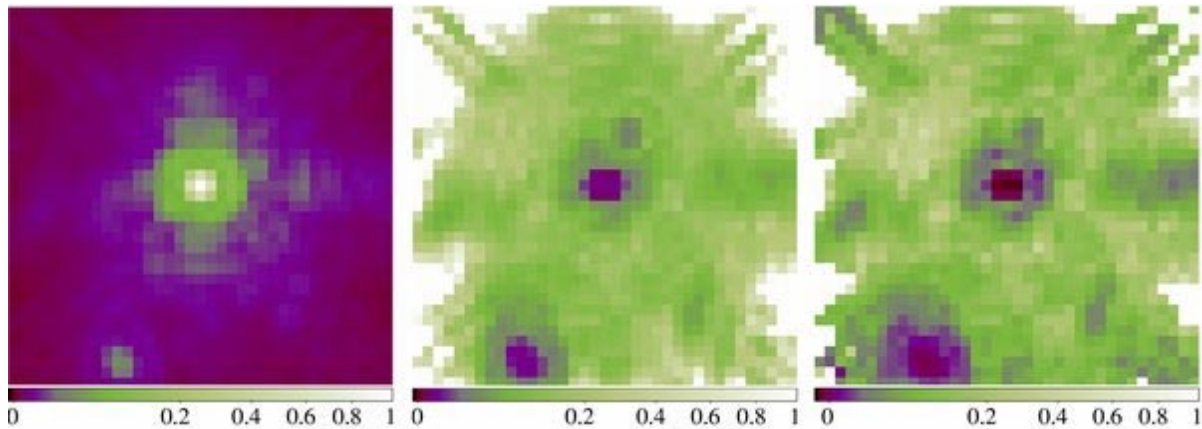

Figure 5. Simulated triple-star system with the faint companion (Δ*m* = 4.25, separation = 0″.41) located on top of a static speckle just above the first Airy ring, the same order and color scale as in Figure 4.

The new technique has the advantage of simplicity over the previous algorithm (Gladysz & Christou 2008). The computation is almost instantaneous. The disadvantage is that, in its current qualitative form, it does not provide the PFA. This problem will be addressed in Section 3.3. Next, we look at some real double stars to see how the transformations work when the companion's peak is not positioned on a single pixel.

3.2. Real Double Stars

Only a small fraction of our observations were aimed at double stars, and only a few of them turned out to be challenging enough to try new approaches to detection. Here, we present three of the most interesting cases.

3.2.1. HD 8799

Figure 6 shows the SAA image of the binary star ω And (HD 8799, $m_V$ = 4.8, $m_K$ = 3.9, spectral type F5). The only difference in the observational setup between this data set and HD 216756 is exposure time; here it was set to 22 ms. The StarFinder PSF-fitting algorithm (Diolaiti et al. 2000) was used to get an a priori estimate of relative photometry and astrometry of the system. This code was designed for the analysis of crowded fields imaged with AO and the fitting algorithm can take advantage of many estimates of the PSF within the field of view. In our work, we provide StarFinder with the PSF estimate – a single star matched as closely as possible to the science object. In the case of a double star we only use a subset of StarFinder's capabilities, while still obtaining valid results.

As calibration PSF we used the SAA image of the star SAO 55171. This star was observed almost an hour before the science target and its spectral properties do not match HD 8799 (see the discussion in Gladysz & Christou 2008). Therefore, StarFinder's results have to be treated as approximate even though the shape of the two PSFs was qualitatively very similar. The measured magnitude difference and separation were $\Delta m_K$ = 3.64 and θ = 0″.667, respectively.

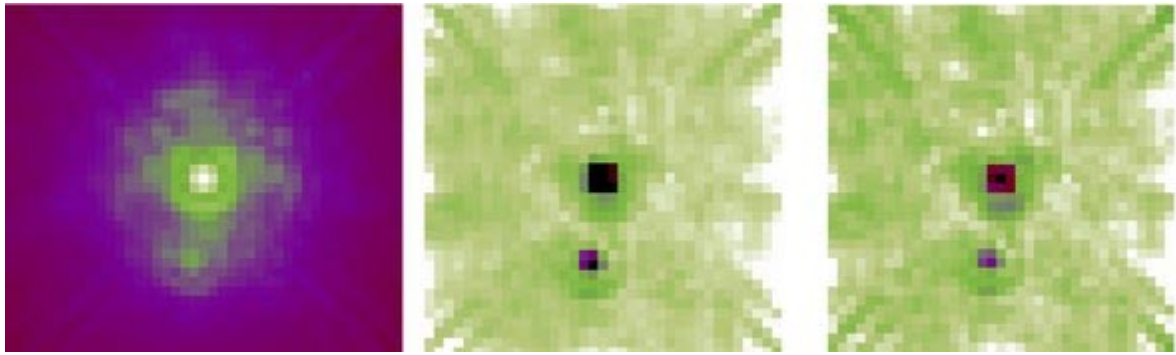

Figure 6. Binary star HD 8799: real SAA image (left), $I_s/I_c$ (center), $m_3 \cdot I_s/I_c$ (right), the same color scale as in Figure 4.

Again, the statistical maps smoothed out the anisotropies that were present in the SAA image. Note how flat the central map is, with the exception of spiders. The faint companion is much more visible in the central and right panels as compared to the SAA image.

3.2.2. HD 235089

For this star ($m_V = 8.8$, $m_K = 6.41$, spectral type K5) we had a very good calibration PSF: HD 235160 ($m_V = 9.16$, $m_K = 6.52$, spectral type K0) observed only 10 minutes later with very stable seeing. The zenith angle for both observations was almost identical and equal to 20°. The exposure time was set to 57 ms. Results from StarFinder were $\Delta m_K = 3.98$ and $\theta = 0\rlap{.}''585$. Figure 7 shows the application of Equations (10) and (11) to this data set.

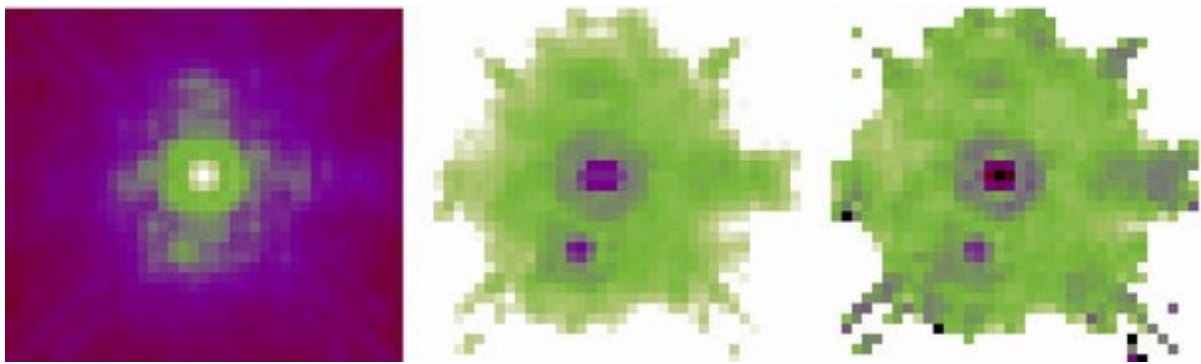

Figure 7. Binary star HD 235089, the same order and color scale as in Figure 6.

The $I_s/I_c$ map is again very smooth. It should be noted that the $m_3 \cdot I_s/I_c$ map (the rightmost panel in Figure 7) is more prone to false alarms as evidenced by the black and purple pixels, but these are isolated and scattered in contrast to the effect induced by the real source.

3.2.3. HD 170648

For HD 170648 ($m_V = 8.04$, $m_K = 7.21$, spectral type A2) we observed a properly matched calibration PSF (HD 173869: $m_V = 7.9$, $m_K = 7.53$, spectral type A0). Again, the temporal gap between these observations was no longer than 10 minutes, and zenith angle in both cases was 15°. HD 170648 had significantly lower S/N than the stars discussed before because it is fainter and because the exposure time was set to 22 ms. This meant that many pixels had simple Gaussian statistics and therefore the statistical maps (Figure 8) are limited spatially. For this system StarFinder yielded $\Delta m_K = 3.07$ and $\theta = 0\rlap{.}''687$.

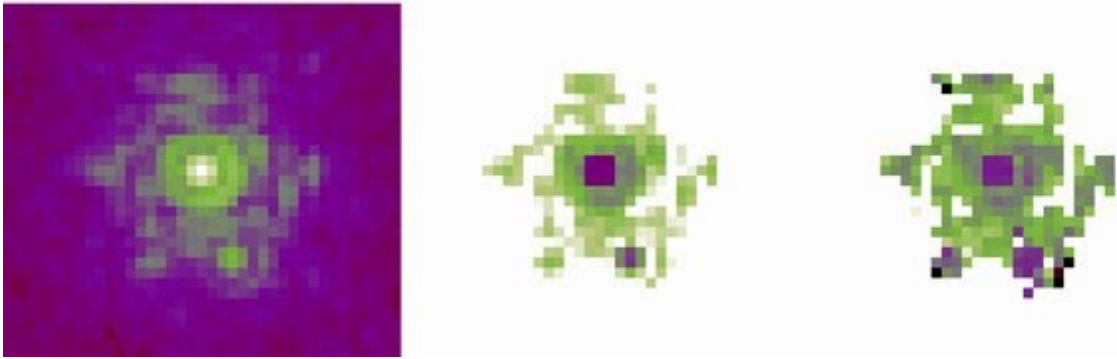

Figure 8. Binary star HD 170648, the same order and color scale as in Figure 6.

Again, there appear false alarms in the $m_3 \cdot I_s/I_c$ map, i.e., some speckle pixels have very low values after the transformation. This is the consequence of low flux from the star. The statistics of speckle and core intensity get lost in readout and background noise. The estimation of $I_s$, $I_c$, and skewness is therefore unreliable. Especially the skewness sample estimator – Equation (12) – is susceptible to noise. We emphasize that these false alarms are isolated as opposed to the trace of a true detection coming from *all* the pixels containing the companion's peak.

The conclusion from these tests is that the $I_s/I_c$ map is more reliable than $m_3 \cdot I_s/I_c$. We therefore adopt Equation (10) for further studies.

3.3. Dynamic Range

Given the isotropic character of the $I_s/I_c$ map (e.g., in Figures 4-6) it is natural to compare the map's value at the location of a suspected companion to the histogram of the $I_s/I_c$ speckle values, which are randomly distributed about a common mean value and we postulate that they come from the same distribution. This is the simplest, but not necessarily the most accurate, approach to the computation of the PFA. Its accuracy is limited by the finite character of the speckle sample, which means one only has access to the estimate of the distribution, i.e., the histogram. The PDF of the $I_s/I_c$ test statistic is not known. We use the term "test statistic" because in this section we will be discussing issues pertaining to the discipline of hypothesis testing. The observer (the detection algorithm) computes the value of the test statistic ($I_s/I_c$) for every location in the image. The value obtained for the suspected source is then compared to the histogram of all the other values. The observer can then accept the hypothesis that the companion is actually present in the image if the measured value is lower than a predefined threshold. The number of speckle pixels with values of $I_s/I_c$ lower than the investigated pixel translates to the PFA.

The top panel of Figure 9 shows the histogram of the $I_s/I_c$ values for the star HD 216756, shown in the left panel of Figure 3. The histogram was computed only for the speckle pixels, i.e., pixels that were dominated by speckle, as opposed to readout and background noise. The two sources were masked so only speckle statistics are present in the histogram. The two vertical lines on the left of the histogram correspond to two artificial companions which were positioned on the zero of the speckle distribution as in Figure 4: one was the same as before ($\Delta m = 4.25$), while the second was made much fainter ($\Delta m = 6.1$). Over-plotted is the $I_s/I_c$ histogram for the star HD 235089 (shown in Figure 7). In the central and bottom panels of Figure 9 we show the $I_s/I_c$ histograms for stars HD 8799 and HD 170648.

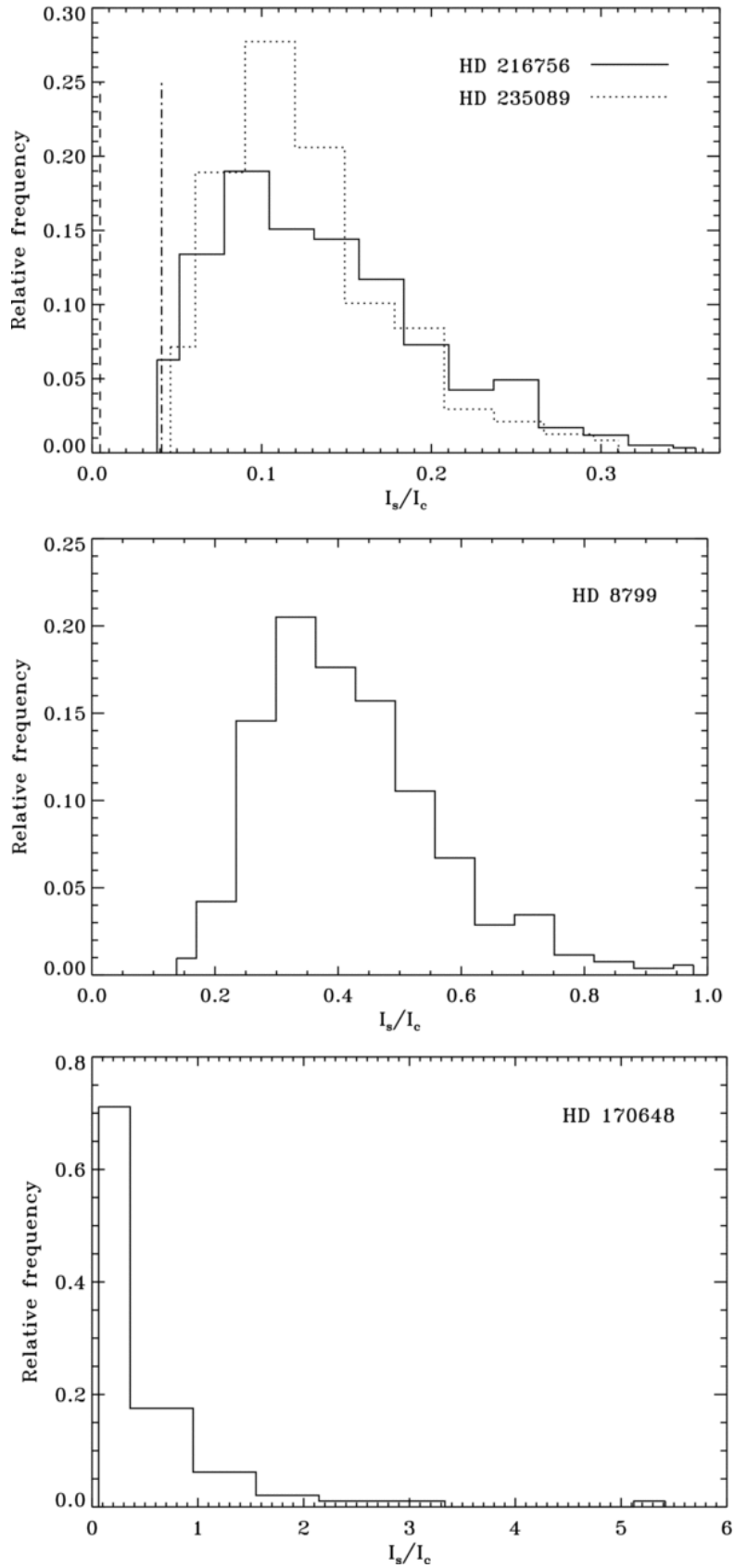

Figure 9. Top: histograms of the $I_s/I_c$ values for the stars HD 216756 (solid line), and HD 235089 (dotted line). The two vertical lines on the left of the histogram correspond to two artificial companions positioned 5 pixel down and 3 pixel left of the center: $\Delta m = 4.25$ (dashed line), and $\Delta m = 6.1$ (dash-dot line). Center: $I_s/I_c$ histogram for the star HD 8799. Bottom: $I_s/I_c$ histogram for HD 170648. All histograms were normalized, i.e., divided by the number of speckle pixels.

The first two histograms are displayed on the same plot because they are very similar in shape and values. Nevertheless, the plots for stars HD 8799 and HD 170648 indicate that $I_s/I_c$ can take on any value, and the underlying PDF can change shape. The HD 170648 histogram is an interesting case: the plot is highly positively skewed and suggests the exponential distribution. Here, $I_s/I_c$ reached high values because of the extra intensity variability induced by the relatively high readout and background contributions (this is a low-flux data set—see the discussion in Section 3.2.3). We leave for the future analytical work on the $I_s/I_c$ PDF noting that the robust gamma distribution is going to be the first candidate (Gladysz et al. 2008a).

The $I_s/I_c$ histogram forms the basis for the detection hypothesis. The lack of an analytic distribution for the ratio $I_s/I_c$ means that for high-S/N companions one cannot compute the exact value of the PFA. The PFA estimate can only be computed when some speckles have lower values of $I_s/I_c$ than the companion (low-S/N case). This is illustrated in the top panel of Figure 9. The dashed vertical line corresponding to the $\Delta m = 4.25$ companion is far from the $I_s/I_c$ histogram, i.e., there were no speckles with similarly low values of the $I_s/I_c$ ratio. In this case, an analytic distribution would have to be fitted to the histogram in order to find the PFA. For the $\Delta m = 6.1$ companion (dash-dot line) there was one speckle pixel with lower $I_s/I_c$. The total number of speckle pixels was 590, and so the PFA estimate for this case is $1/590 = 0.0017$. With a small sample like this, the estimation of PFAs of the order of $10^{-2}$ or lower is very difficult.

With this limitation in mind, we approach the problem of finding the dynamic range of the new method using data-based distributions. The S/N of $3\sigma$ was used. Assuming Gaussian statistics this translates to the PFA of 0.00134—this is the integrated power in the Gaussian PDF starting at $3\sigma$ above the mean (Gladysz & Christou 2008). Faint companions of varying relative magnitudes were inserted in the data set HD 216756. Three separations were explored: $0\rlap{.}''45$, $0\rlap{.}''75$, and $1''$. For each of these separations eight locations were investigated to check the variability of the method due to positioning of the companion on and off the static speckles. The detection threshold was determined as the magnitude difference that resulted in precisely 1 or 2 pixels having an $I_s/I_c$ value lower than that at the position of the fake companion. Given the total number of pixels, this corresponds to a PFA of 0.0017-0.0034, similar to $3\sigma$ for a Gaussian distribution. The resulting dynamic range plot is shown in Figure 10.

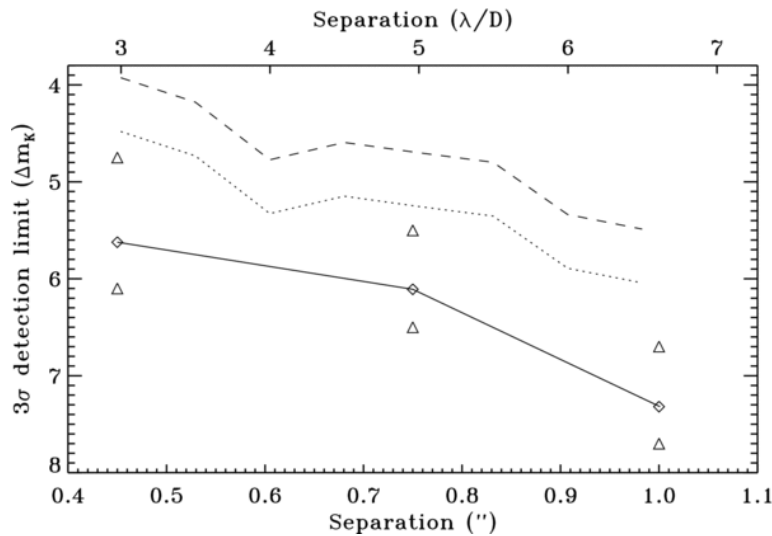

Figure 10. Dynamic range plot for the Lick Observatory AO system coupled with the short-exposure mode of the IRCAL camera. The diamonds represent the mean values for the eight locations investigated for each separation. The triangles denote the maximum and minimum $3\sigma$ detections obtained for these separations. Overplotted are the $3\sigma$ (dotted line) and $5\sigma$ (dashed line) detection limits for the SAA image of HD 216756.

The system dynamic range is usually calculated with an implicit assumption of spatial stationarity, i.e., common statistics for pixels located in annuli centered on the origin. This conventional approach also assumes common Gaussian confidence levels, which is inaccurate for anisotropic PSFs, as explained in Section 3.1. Nevertheless, we plot the dynamic range for "normal" SAA imaging in order to show a rough estimate of the improvement obtained with the new method. For this we assume that the companion's signal must be higher than $3\sigma$ or $5\sigma$ of intensity in the SAA image of HD 216756. Standard deviations were computed in annuli centered on the origin. The $3\sigma$ and $5\sigma$ values were divided by the SAA peak intensity and this brightness ratio was converted to magnitude difference. It is interesting to note how the two static speckles located at separations around $0''\!\!.7$ right and above the PSF core (see Figure 3) influence these two plots. It has been suggested to use the more conservative $5\sigma$ criterion to avoid false alarms resulting from non-Gaussian spatial statistics in AO data (Kasper et al. 2007; Marois et al. 2008). We obtain almost a 2 mag gain in achievable contrast by using our method compared to $5\sigma$ detections in SAA images. The gain relative to the $3\sigma$ dynamic range is a respectable 1.5 mag.

4. DIFFERENTIAL ASTROMETRY

It is straightforward to extend the detection algorithm to find the relative position of a faint companion. In order to estimate the subpixel location one simply needs to interpolate the data cube around the unknown position of a suspected source. The peak of the companion's PSF coincides with the minimum of this interpolated map. After identification of a source based on the approach described in Section 3, and subsequent determination of its center, one can interpolate the data cube around the position of the companion and extract the time series corresponding to its precise subpixel location. This time series can then be used for other purposes, such as differential photometry as outlined in Section 5.

We used the IDL function INTERPOLATE that implements the cubic convolution interpolation method (Park & Schowengerdt 1983). Cubic convolution determines the value at the specified location from the weighted average of the 16 neighboring pixels. The algorithm closely approximates the theoretically optimum sinc interpolation. The 5×5 neighborhood of the companion's peak was subsampled with the spatial interval of 0.25 pixel. The ratio $I_s/I_c$ was then computed for each interpolated time series. These values were then inverted for better visual effect (left panel in Figure 11). The location of the maximum was often very close to the true position of the companion (less than a pixel away in most cases), but for further improvement in accuracy a two-dimensional function was fitted to the data. For simulated companions located further than $0''\!\!.5$ from the stars the interpolated $I_s/I_c$ map had a flat top around the maximum (left panel in Figure 11). Initially, we employed a two-dimensional Gaussian function for the fitting process, but the results were not satisfactory and the following Butterworth function was used instead:

$$f(\mathbf{r}) = a + \frac{b}{\sqrt{1 + \left(\frac{\mathbf{r}}{f_C}\right)^{2N}}}, \quad (13)$$

where $a$ and $b$ are the fitting parameters which correspond to the base and scale values, $\mathbf{r}$ is the position vector in the focal plane, $f_C$ is the cutoff frequency (this function is primarily used in Fourier-filtering of images), and $N$ is the order of the function. We used the excellent MPFIT package for function fitting. Figure 11 illustrates how well the Butterworth function approximates the $I_s/I_c$ profile in the vicinity of the companion's peak. This case corresponds to $\Delta m = 4.25$, $\theta = 1''\!\!.2$. The Gaussian gave better results than the Butterworth function for very small separations (less than $0''\!\!.5$), for which only a small area (3×3 pixels) was interpolated to avoid inclusion of the Airy rings. In these situations, the interpolated $I_s/I_c$ map had a more Gaussian-like shape.

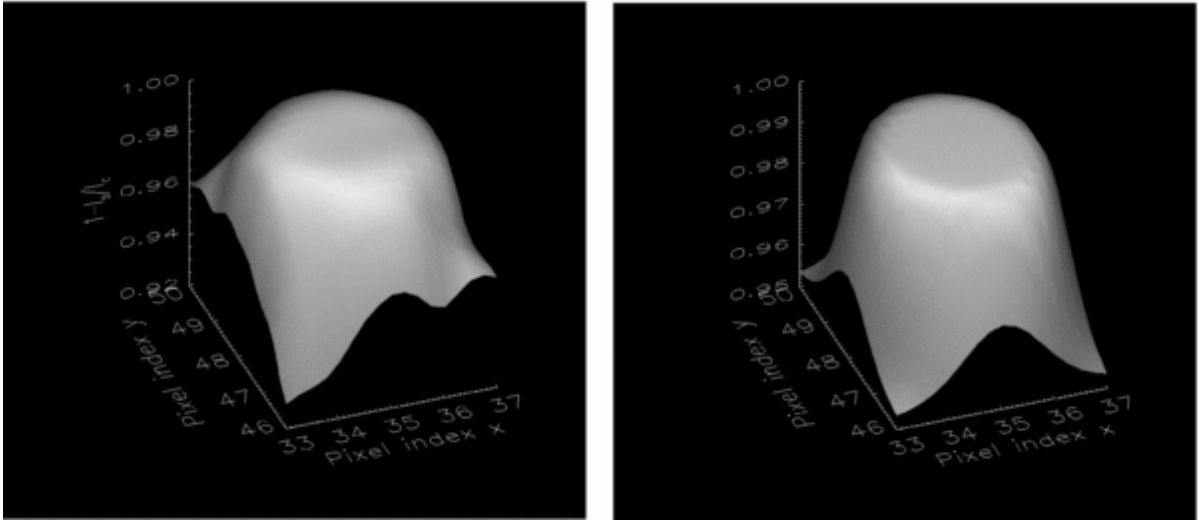

Figure 11. Illustration of the SSD relative astrometry performed on an artificial companion (separation = $1\rlap{.}''2$, $\Delta m = 4.25$) inserted in the HD 216756 data set. Left: the interpolated neighborhood of the companion's peak pixel converted to a map of $1-I_s/I_c$. Right: two-dimensional Butterworth function fitted to the $1-I_s/I_c$ map. The astrometric error is $0\rlap{.}''002$ here.

In this section, we describe the astrometric experiments on simulated companions inserted in the HD 216756 data set. The accuracy of differential astrometry was tested for three separations and differential magnitudes ranging from 2.5 to 5.3. Investigated separations were $0\rlap{.}''5$, $0\rlap{.}''75$, and $1\rlap{.}''1$; for each separation eight positions for the companion were tested in order to minimize the bias from anisotropic PSF. The mean absolute astrometric error was computed based on the results from these eight positions. The results are presented in Figure 12. We want to stress that this is a preliminary analysis and Figure 12 should be viewed as a "first guide" to the accuracy of our method which can be improved by optimizing the fitting process for a particular location and differential magnitude.

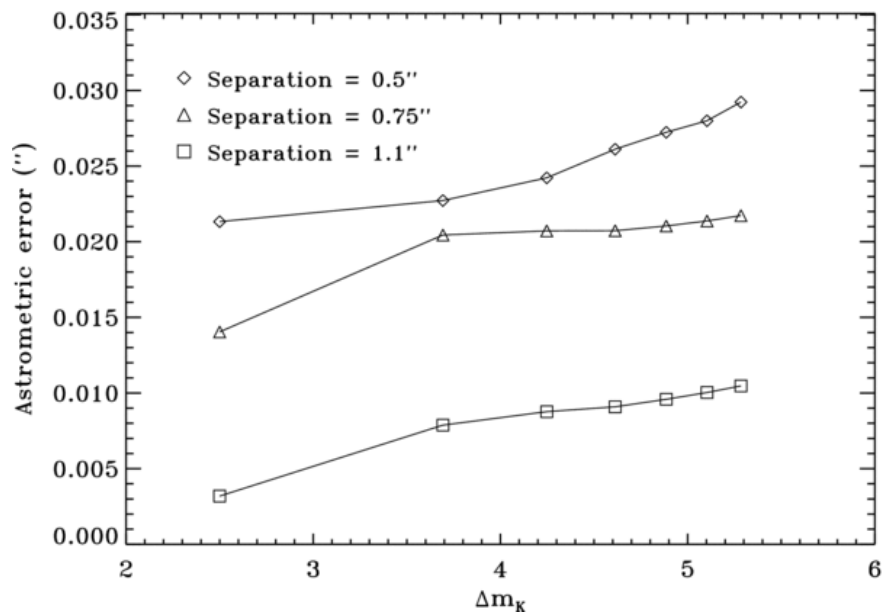

Figure 12. Mean absolute astrometric error for the three separations within the speckle cloud of HD 216756. The error was obtained for simulated companions with $\Delta m = 2.5$ to $\Delta m = 5.3$. Pixel size is $0\rlap{.}''076$.

We observed that the transformation given by Equation (11) gave better results than Equation (10) for small magnitude differences, but completely failed for Δm greater than 4.5 so these results are not shown here. Statistical speckle discrimination yields differential astrometry better than $0\rlap{.}''01$ for separations larger than $1''$. For smaller separations the error can reach $0\rlap{.}''03$, which is still less than half a pixel.

## 5. DIFFERENTIAL PHOTOMETRY

Two random processes take place at the location of the companion's peak (ignoring Gaussian and Poissonian noise). The speckle and signal intensities add, and the PDFs are convolved. The analytical form of these PDFs is known. This offers an incentive to develop a "PDF deconvolution" technique. The distribution of the signal (peak intensity) described by Equation (8) is "blurred" by the speckle kernel given by Equation (1). Between them, these distributions have six parameters. Fortunately, the first three parameters of the signal's PDF are common for the bright star and the companion. These parameters could be first estimated for the bright star, so that the deconvolution problem is constrained to finding only three parameters: $I_c$, $I_s$, and $I^*_{peak}$. We show how this could be done by minimizing the difference between an estimate of the composite PDF and the observed histogram.

### 5.1. The "PDF Deconvolution" Approach

In what follows we used the data set corresponding to the star HD 216756. First, we looked at the fit of Equation (8) to the bright-star histogram. The least-squares, one-dimensional fitting routine MPFITFUN was used. The parameters we wanted to estimate were $k$, $\theta$, and $\mu$ (see the text below Equation (5)). The term to be minimized is

$$E = \sum_{j=0}^{n-1} [h_j - \hat{p}_{on}(I_j; k, \theta, \mu, I^*_{peak})]^2, \qquad (14)$$

where $h$ is the histogram with $n$ bins, and $\hat{p}_{on}(I_j; k, \theta, \mu, I^*_{peak})$ is the estimate of the on-axis PDF described by Equation (8) evaluated for intensities corresponding to the histogram bins $I_j$.

$I^*_{peak}$ is the peak value of the diffraction-limited image normalized to have the same power as the observed star. It is also estimated in the process and its value will be used for differential photometry as will be discussed at the end of this section. Nevertheless, we will now devote two paragraphs to showing how $I^*_{peak}$ obtained from the histogram fit can also be used to obtain the Strehl ratio. In differential photometry, knowledge of SR might be helpful when several data sets are concerned. Then an estimate of SR for the preceding observations can constrain $I^*_{peak}$ for a given star, making the fit described by Equation (14) more reliable.

The Strehl ratio can be computed as the mean value of the bright star's peak time series divided by $I^*_{peak}$ (Equation (6)). With binary stars estimation of SR with long exposures is very inaccurate when the companion is within the speckle cloud because the normalization implied in Equation (6) cannot be carried out (the total flux of the central star is not known). We can therefore obtain estimates of SR from one-dimensional fitting in the cases when standard SR calculation is impossible. This is a direct extension of recent work published by Soummer & Ferrari (2007) on SR estimation. Figure 13 shows the fit of Equation (8) to the bright-star histogram for HD 216756.

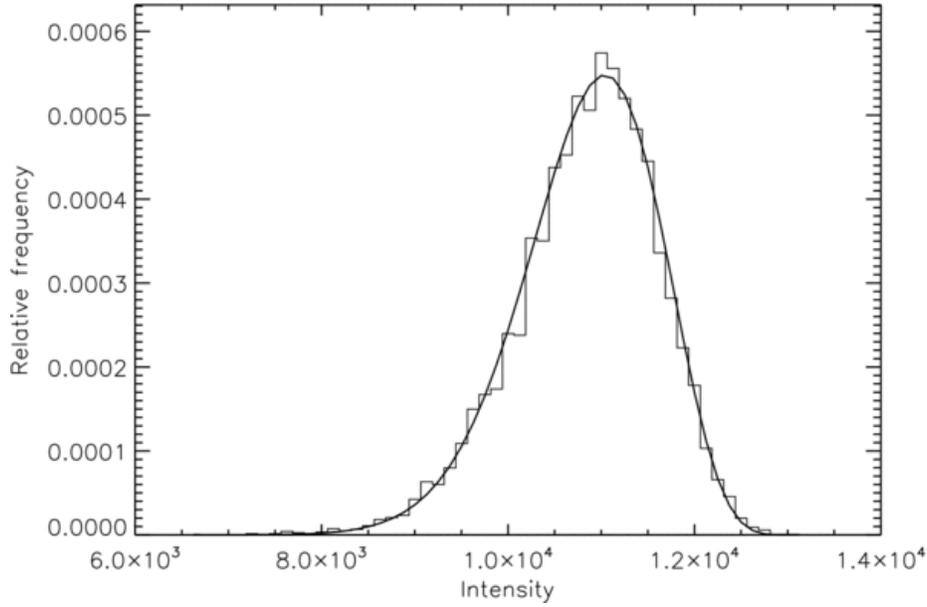

Figure 13. Least-squares fit of Equation (8) to the intensity histogram at the central point in HD 216756.

For the fitting process we assumed bounds for SR (0.3-0.55), and used the Maréchal approximation (Born & Wolf 1980) to relate these limits to the phase variance. The algorithm proved very insensitive to the initial estimates of $k$, $\theta$, and $\mu$ and always yielded the same answer. The value of $I^*_{peak}$ obtained from the fit was plugged into Equation (6) together with the mean peak intensity. This way we obtained an independent estimate of SR equal to 0.51. This is precisely the value we got from the standard approach (the companion in the bottom left corner was outside the flux-normalization area).

As stated before the values for $k$, $\theta$, and $\mu$ are common for both the central star and the companion. In the next fitting process – at the location of the companion – these parameters were kept fixed, and the algorithm only searched for $I_c$, $I_s$, and $I^*_{peak}$ which provided the best fit to the companion's histogram. The term to be minimized is now

$$E = \sum_{j=0}^{n-1} [h_j - \hat{p}_{on}(I_j; k, \theta, \mu, I^*_{peak}) \otimes \hat{p}_{off}(I_j; I_c, I_s)]^2,$$

(15)

where $\hat{p}_{off}(I_j; I_c, I_s)$ is the estimate of the off-axis PDF given by Equation (1), and $k$, $\theta$, and $\mu$ are now kept fixed. $\otimes$ denotes convolution.

Several issues concerning the fitting process were explored. First, the influence of the histogram bin width was tested. We used optimal bin width (Scott 1979) for histograms with skewness close to zero. We also applied a correction to the optimal bin size when histograms displayed positive skewness, but this had no effect on the results (photometric errors). We examined the effect of various weighting functions on the fit accuracy, and again this had negligible effect. Finally, we checked the influence of subtracting trends (via linear regression) from the time series, but no significant differences were observed.

We observed that the algorithm was robust and proved insensitive to large changes in initial estimates for $I_c$ and $I_s$. As long as the sum of $I_c$ and $I_s$ was the same order of magnitude as the mean intensity at a given location, the algorithm had no problem converging.

In Figure 14, we plot the intensity histogram at the location of the companion from Figure 4, together with the constrained fit of convolution of Equations (8) and (1). The histogram was divided by its integrated power to obtain an estimate of the composite PDF.

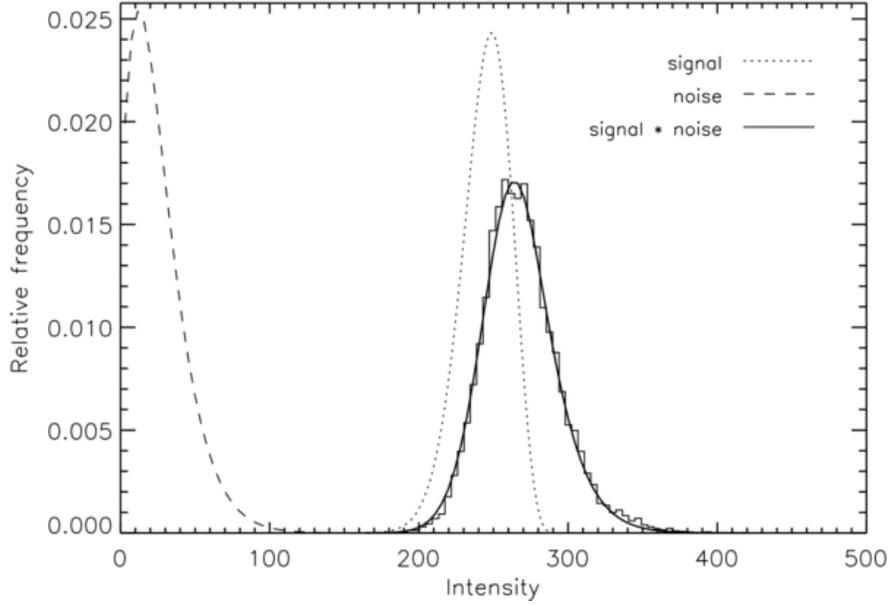

Figure 14. Result of "PDF deconvolution" for the case depicted in Figure 4 ($\Delta m = 4.25$, separation $= 0\farcs 45$). "Signal" is the on-axis PDF $p_{\rm on}$, "noise" is the off-axis speckle PDF $p_{\rm off}$, and * denotes convolution.

The estimated signal PDF $\hat{p}_{\rm on}(I_j;k,\theta,\mu,I^*_{\rm peak})$ can be numerically integrated to get the signal's mean value,

$$E(I) = \int_{-\infty}^{\infty} I \cdot p(I) dI. \quad (16)$$

Using SR, this mean peak intensity could be converted to an estimate of the total flux of the companion. In the future we will test this approach to absolute photometry. In this paper, we are concerned with differential photometry, which can be obtained by relating $I^*_{\rm peak}$ obtained in Equation (15) to $I^*_{\rm peak}$ from Equation (14). Because this parameter contains information about the object's flux, the ratio

$$\widetilde{\Delta m} = -2.5 \log_{10} \left( \frac{I^*_{\rm peak}({\rm companion})}{I^*_{\rm peak}({\rm star})} \right) \quad (17)$$

is a very good estimator of differential photometry. In our tests this value was then compared to the known contrast of the fake companion:

$$\text{photometric error} = \Delta m - \widetilde{\Delta m}, \quad (18)$$

where $\Delta m$ is the true magnitude difference, and $\widetilde{\Delta m}$ is its estimate.

In order to validate the "PDF deconvolution" approach we have compared it with photometry obtained by PSF-fitting. We used the StarFinder algorithm with calibration PSFs, as described below. It should be stated that packages have been developed for the specific task of AO astrometry and photometry of binary stars (Roberts et al. 2005; Burke et al. 2008), so the results quoted here may not be state-of-the-art in this discipline. Nevertheless, the StarFinder results are representative of the class of PSF-fitting algorithms.

5.2. Accuracy of "PDF Deconvolution" Versus StarFinder for Simulated Companions

"PDF deconvolution" was tested using the HD 216756 observations. For StarFinder we needed two data sets – one to simulate double stars, and the other as a calibration PSF. We did not have such matched data sets with SR similar to that of HD 216756 (SR = 0.51). Instead we used three data sets corresponding to the same star (HD 153832 $m_V = 7.25$, $m_K = 4.78$, spectral type K0), observed continuously for 1 hr. The three data sets were observed one after the other but due to variable seeing Strehl ratios of the SAA images were 0.49, 0.43, and 0.42. These SAA images formed the input for StarFinder, and we simulated two general cases: properly matched PSFs (0.43 and 0.42 SR pair), and mismatched PSFs (0.49 and 0.42 SR pair).

To make the comparison between StarFinder and "PDF deconvolution" fair, we used our method "blindly" in terms of companion location. This location was estimated using the approach to astrometry outlined in Section 4. StarFinder had to be given guesses for the companion's position and we observed that these guesses had to be exact; otherwise the algorithm moved in a different direction and converged on the brightest static speckle.

In Figure 15, we present the comparison between our method and StarFinder for the scenario when the companion was located on the intensity minimum (shown in Figure 4), as well as when it was positioned on a static speckle (Figure 5). The intensity ratio between the star and the companion was varied between 10 and 250. Differential magnitude was computed with the help of Equation (17); photometric error was computed via Equation (18).

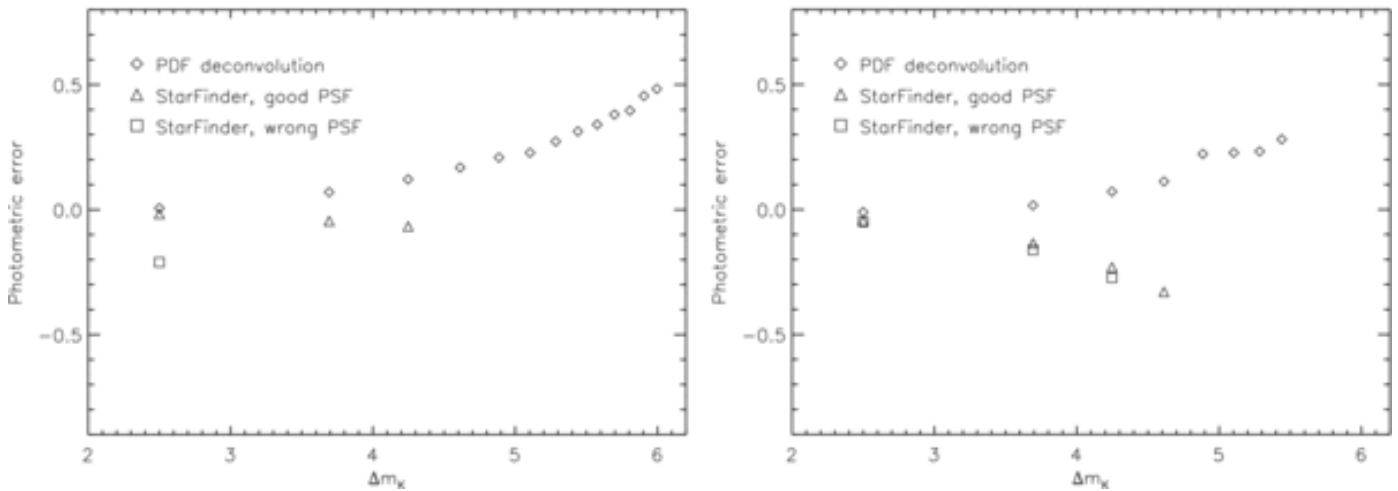

Figure 15. Comparison between "PDF deconvolution" approach to photometry and StarFinder. Two scenarios of properly matched PSFs, and mismatched PSFs are shown for StarFinder. Left: companion located on the zero of the speckle distribution, as in Figure 4. Right: companion positioned on a static speckle, as in Figure 5.

In Figure 15, we only show the results when StarFinder converged on the location of the companion or the neighboring pixels. As can be seen this only happened 11 times for these high-dynamic-range systems. For the more fortunate conditions (companion positioned on intensity zero) PDF deconvolution yields results comparable to StarFinder's with the correct PSF down to $\Delta m = 4.25$. For larger magnitudes StarFinder does not converge while our method still gives accurate results (error between 0.1 and 0.5). When the companion is located on a static speckle the performance of PDF deconvolution deteriorates compared to the first scenario. For $\Delta m$ larger than 4.45 the technique yields invalid results due to large errors in astrometry. In both scenarios the new method kept the photometric error below 0.25 down to $\Delta m = 5$.

In the preceding analysis, we only used a single location for the companion in order to emphasize the dependence of the new method on the strength of the underlying speckle variance. We will now present the performance of our deconvolution method as a function of separation. Similarly to the astrometric experiments companions were positioned at eight locations, which had a common separation from the bright star. This way PSF anisotropy was accounted for in the tests. Below we plot mean absolute errors.

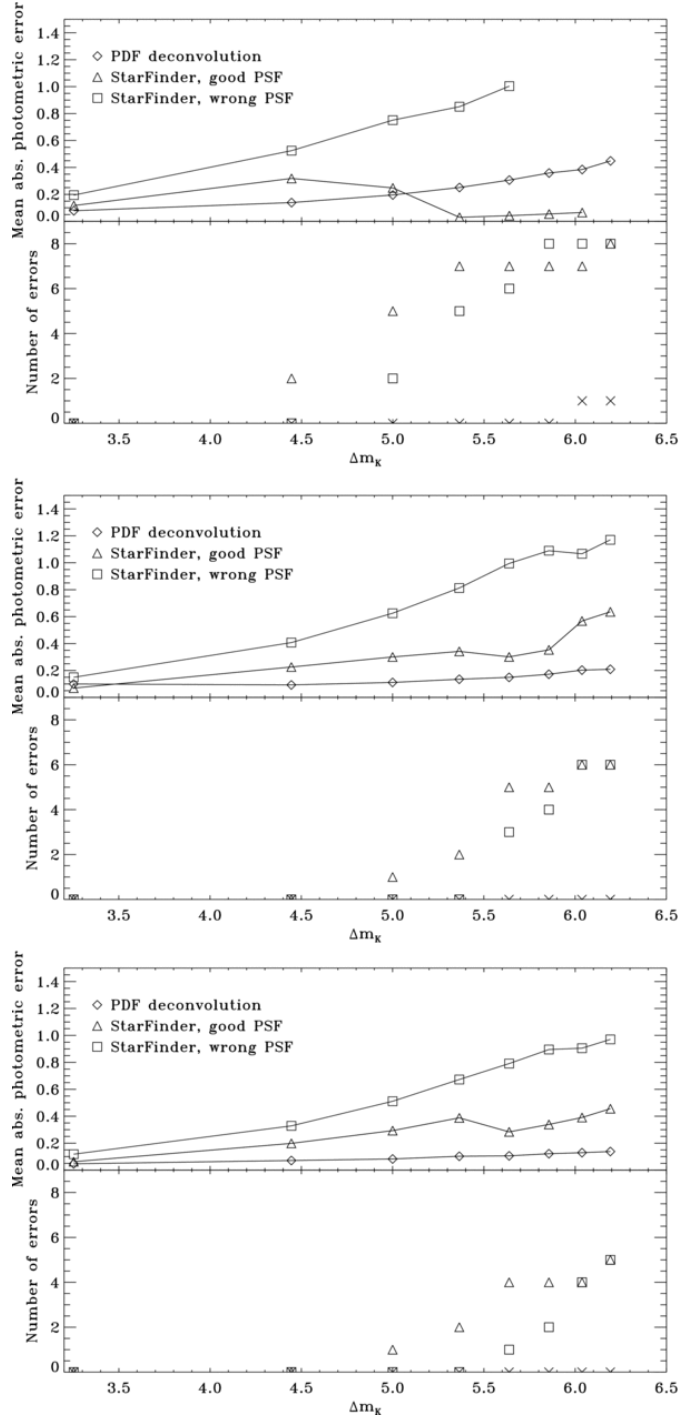

Figure 16. Comparison between "PDF deconvolution" approach to photometry and StarFinder, separations = 0."6 (top), 0."75 (middle), and 0."9 (bottom). Mean absolute photometric error obtained for a sample of eight locations is plotted in the top panels. The number of errors is shown in the bottom panels.

In Figure 16, the number of errors is also plotted. An error was an instance when StarFinder did not converge, or converged on a static speckle. In the case of "PDF deconvolution," error was a situation when histogram-fitting process failed. The new method is much more reliable than PSF-fitting. There were no errors for "PDF deconvolution" for almost all simulated contrasts, as opposed to four or more for StarFinder when used in the most challenging scenarios. For the smallest tested separation of $0''\!.6$ PSF-fitting is highly unreliable for $\Delta m > 5$. Given a mismatched PSF the algorithm did not converge once for $\Delta m > 5.7$. Provided with a good calibration PSF it did not converge seven out of eight times for magnitude differences larger than 5, hence small photometric errors reported in the top plot are not representative of its performance. "PDF deconvolution" provides excellent reliability, small photometric errors, it is not susceptible to initial guesses, and it is self-calibrating by design. For larger separations of $0''\!.75$ and $0''\!.9$ our method keeps the photometric error below 0.2 mag, less than half of the error yielded by PSF-fitting. These results support the conclusion that "PDF deconvolution" is an excellent tool for AO differential photometry of faint companions.

## 5.3. Real Double Stars

In this section, we present relative astrometry and photometry computed using the new methods for real binary stars described in Section 3.2. For comparison the StarFinder results are quoted again.

### 5.3.1. HD 8799

With our new astrometric technique outlined in Section 4 we obtained a new estimate of the companion's position, which differed from the StarFinder estimate by less than 0.1 pixel. This position translates to a new separation estimate of $0''\!.669$, which is very close to the StarFinder result ($\theta = 0''\!.667$). For "PDF deconvolution" we used the original short exposures binned into groups of $k$ samples, $k = 2, 3,...,10$, and summed. The algorithm was executed on the original data cube and the binned data. The result was the mean $\Delta m_K$ and its standard deviation. We decided to use the bin size as the free parameter because in reality there is always a trade-off between observing with short exposures (which preserve the two distinct PDFs in the images), and longer exposures, which give better S/N of the companion in each frame. For HD 8799 we obtained $\Delta m_K = 3.65 \pm 0.03$, close to the StarFinder's value $\Delta m_K = 3.64$. In Figure 17, we show the result of "PDF deconvolution" for the original short exposures.

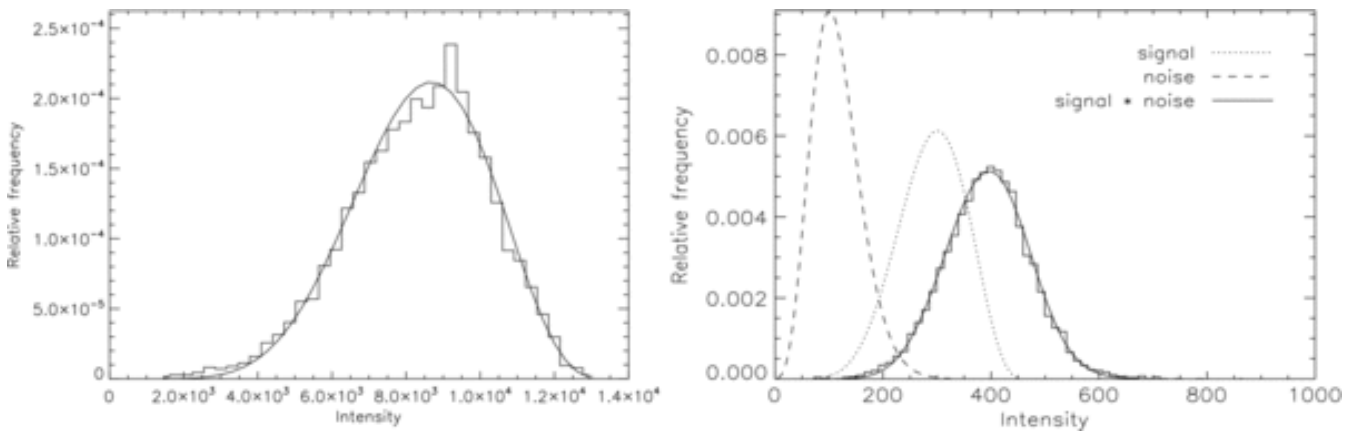

Figure 17. Result of "PDF deconvolution" for the star HD 8799. Left: least-squares fit of the on-axis PDF $p_{on}$ to the bright star's histogram. Right: fitted composite PDF at the subpixel location of the companion together with the signal and speckle PDFs: $p_{on}$ and $p_{off}$.

### 5.3.2. HD 235089

For HD 235089 StarFinder returned $\Delta m_K = 3.98$ and $\theta = 0\farcs585$. Here, the subpixel position of the companion obtained with our astrometric method was 0.25 pixel away from the StarFinder result but the separation value of $0\farcs587$ agreed very closely with the previous estimate. "PDF deconvolution" on the binned data returned six very poor fits. The differential magnitude based on the four reliable outcomes was $\Delta m_K = 4.02 \pm 0.1$. Figure 18 shows the fitted PDFs for the original 57 ms exposures.

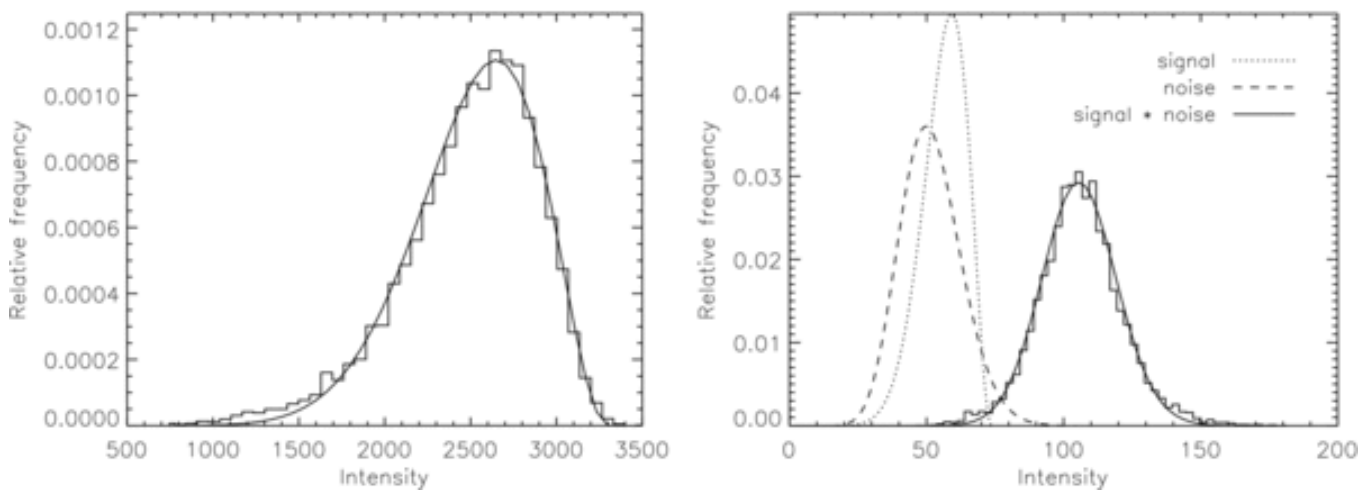

Figure 18. Result of "PDF deconvolution" for the star HD 235089, the same order as in Figure 17.

### 5.3.3. HD 170648

For this system (StarFinder: $\Delta m_K = 3.07$ and $\theta = 0\farcs687$) the observations had to be binned into groups of three frames in order to increase the speckle-to-background noise ratio (see Figure 8). This way a larger $I_s/I_c$ map was obtained and subpixel astrometry could be computed within the neighborhood of the companion's peak pixel. The position we obtained was 0.05 pixel away from the position given by StarFinder. This new position corresponds to a separation estimate of $0\farcs688$. "PDF deconvolution" was executed on the binned data. The result was $\Delta m_K = 2.96 \pm 0.02$. Figure 19 shows the fitted PDFs for the 22 ms data binned into groups of three frames and averaged.

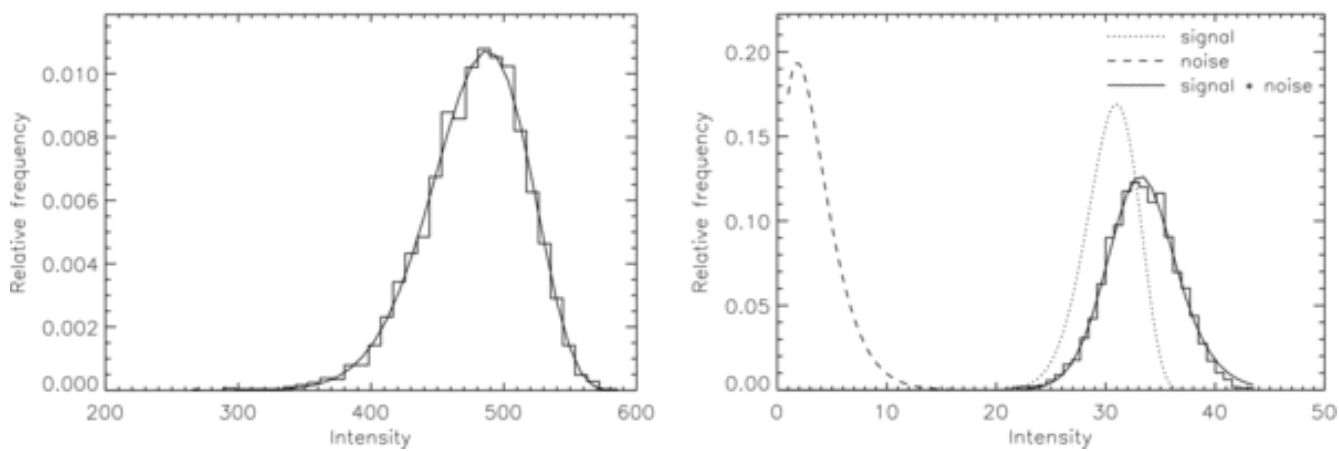

Figure 19. Result of "PDF deconvolution" for the star HD 170648, the same order as in Figure 17.

## 6. SUMMARY

We propose a new class of postprocessing techniques for the detection, astrometry, and photometry of faint companions, and possibly giant exoplanets, with AO. All three algorithms perform very well, even though they were applied to images from a noisy camera. The biggest advantage of the new framework is that no calibration PSF is needed.

Regarding the numbers of short exposures needed for these methods to work, we predict that for detection and astrometry 100 samples will be enough to estimate the scalar $I_s/I_c$ accurately, because the *estimation* error for mean and variance in Equation (9) will be low for 100 samples. Another issue is the longest short-exposure time for which the methods still work. We predict that this time can be quite long (perhaps 1 s in *K* band) because of the correlations observed for on-axis intensity (Gladysz et al. 2006). These correlations mean that the central limit theorem (CLT) will smear the statistics slower than expected for independent samples, so the proposed algorithm could be classified as a "multiple-frame," rather than a "short-exposure" technique. The sensitivity of speckle discrimination to short time series and long exposure times will be investigated in our future work. As a preliminary analysis we inserted five artificial companions in the HD 216756 data set and computed $I_s/I_c$ for data binned into simulated 0.57 s and 5.7 s exposures. Figure 20 suggests that increasing exposure time and reducing the data set size has a small effect on the detection capability of the new method.

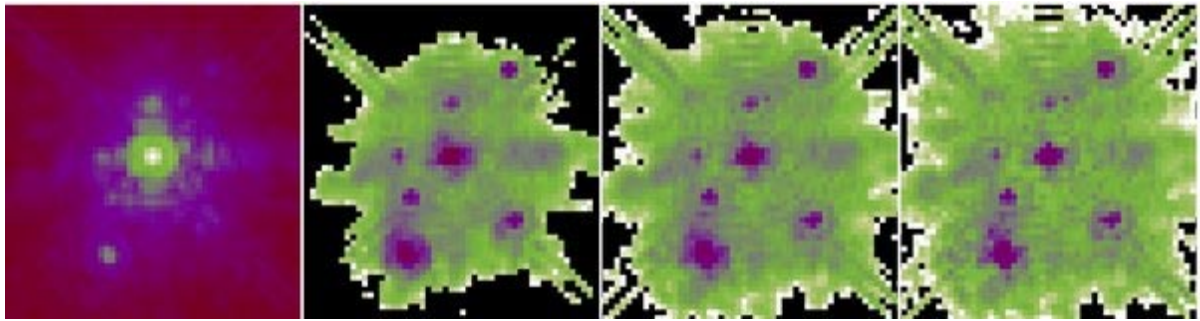

Figure 20. Leftmost panel: SAA image for HD 216756 with one real companion and five simulated companions. Second panel: $I_s/I_c$ map for the original $10^4$ frames with exposure time 57 ms. Third panel: $I_s/I_c$ map for the same data set binned into $10^3$ frames with exposure time 0.57 s. Rightmost panel: $I_s/I_c$ map for the data cube binned into $10^2$ frames with exposure time 5.7 s.

For photometry, an estimate of PDF is needed, i.e., the histogram. The reliability of this PDF estimate will be influenced by the sample size. In this paper, we worked with histograms based on $10^4$ focal-plane measurements. We recently showed that "PDF deconvolution" with 1000 samples already outperforms PSF-fitting (Gladysz et al. 2008b). We expect the performance of this method to deteriorate for samples of less than 100 observations but we leave this analysis for the future. It should also be mentioned that histograms for nonstationary data (highly variable seeing) are worthless as PDF estimates. However, the better the mean correction the more stable it is, so for high-SR systems the problem of nonstationarity will be removed. "PDF deconvolution" requires discernible morphological difference between the on-axis and off-axis intensity PDFs. Based on the Lick Observatory data this occurs at AO compensation with SR > 0.5 although we intend to test this method with low-SR data in the future. Stochastic speckle discrimination does work below 0.5 SR. We have shown that the method yields very good results in *I* band, where AO-compensation delivers images with 0.25 SR on the 5 m telescope at the Palomar Observatory (Gladysz et al. 2008b).

The impact of Poisson noise for very faint sources and short exposures would be to transform all statistics, so that the speckle and companion PDFs will look very similar, invalidating our photometric method. The observer would have to balance that fact against the impact of the CLT, which destroys the PDF asymmetry for longer exposures.

For coronagraphic imaging, where the estimation of the on-axis PDF from the focal plane (Figure 13) is impossible, we propose to use the wavefronts reconstructed from the wavefront sensor to obtain the parameters $k$, $\theta$, and $\mu$. These parameters can be estimated using the method of moments as shown in Gladysz et al. (2008a). We plan to carry out an experiment to test this approach.

In studying exoplanets and substellar companions to bright stars it is almost always necessary to use short exposures to avoid saturation of the detector. We suggest the statistical information that is present in these frames be used in the detection process. It could complement the standard, S/N-based method, which can lead to uncertainties in the determination of the false-alarm probability, especially when the underlying intensity PDF is not taken into account (Marois et al. 2008). We would also encourage implementing short-exposure imaging modes in detectors that do not yet have that capability.

We thank the referee for their careful reading and their suggestions for improvements. This research was supported by Science Foundation Ireland under grants 02/PI.2/039C and 07/IN.1/I906, as well as the National Science Foundation Science and Technology Center for Adaptive Optics, which is managed by the University of California at Santa Cruz under cooperative agreement AST 98-76783. We thank the staff of Lick Observatory, in particular Elinor Gates and Bryant Grigsby. In addition, we thank Michael Fitzgerald for information about the high-speed mode of the IRCAL camera, Nicholas Devaney for useful comments on the draft version of this paper, and Chris Dainty for support for this research.

The authors wish to acknowledge the SFI/HEA Irish Centre for High-End Computing (ICHEC) for the provision of computational facilities and support.

REFERENCES

Beuzit, J.-L. et al. 2006, The Messenger, 125, 29
Biller, B. A., Close, L., Lenzen, R., Brandner, W., McCarthy, D. W., Nielsen, E., & Hartung, M. 2004, Proc. SPIE, 5490, 389
Biller, B. A., Close, L. M., Lenzen, R., Brandner, W., McCarthy, D., Nielsen, E., Kellner, S., & Hartung, M. 2006, IAU Colloq. 200, Direct Imaging of Exoplanets: Science & Techniques, ed. C. Aime & F. Vakili (Cambridge: Cambridge Univ. Press), 571
Born, M. & Wolf, E. 1980, Principles of Optics (New York: Pergamon)
Burke, D., Devaney, N., Gladysz, S., Barrett, H., Whitaker, M., & Caucci, L. 2008, Proc. SPIE, 7015, 70152J
Cagigal, M. P. & Canales, V. F. 1999, J. Opt. Soc. Am. A, 16, 2550
Cavarroc, C., Boccaletti, A., Baudoz, P., Fusco, T., & Rouan, D. 2006, A&A, 447, 397
Diolaiti, E., Bendinelli, O., Bonaccini, D., Close, L., Currie, D., & Parmeggiani, G. 2000, A&AS, 147, 335
Fitzgerald, M. P. & Graham, J. R. 2006, ApJ, 637, 541
Gisler, D. et al. 2004, Proc. SPIE, 5492, 463
Gladysz, S. & Christou, J. C. 2008, ApJ, 684, 1486
Gladysz, S., Christou, J. C., Bradford, L. W., & Roberts, L. C. J. 2008a, PASP, 120, 1132
Gladysz, S., Christou, J., Kenworthy, M., Law, N., & Dekany, R. 2008b, Proc. Advanced Maui Optical and Space Surveillance Technologies Conf. (held in Wailea, Maui, Hawaii, 2008 September 17-19), ed. S. Ryan, (Kihei: The Maui Economic Development Board), E42
Gladysz, S., Christou, J. C., & Redfern, M. 2006, Proc. SPIE, 6272, 62720J
Guyon, O. 2006, IAU Colloq. 200, Direct Imaging of Exoplanets: Science & Techniques, ed. C. Aime & F. Vakili (Cambridge: Cambridge Univ. Press), 559
Hinkley, S. et al. 2007, ApJ, 654, 633
Itoh, Y., Oasa, Y., & Fukagawa, M. 2006, ApJ, 652, 1729
Janson, M., Brandner, W., Henning, T., & Zinnecker, H. 2006, A&A, 453, 609

Kasper, M., Apai, D., Janson, M., & Brandner, W. 2007, A&A, 472, 321
Kasper, M. et al. 2008, Proc. SPIE, 7015, 70151S
Labeyrie, A. 1995, A&A, 298, 544
Lafrenière, D. et al. 2007, ApJ, 670, 1367
Macintosh, B. et al. 2008, Proc. SPIE, 7015, 701518
Marois, C., Doyon, R., Nadeau, D., Racine, R., Riopel, M., Vallée, P., & Lafrenière, D. 2005, PASP, 117, 745
Marois, C., Lafrenière, D., Doyon, R., Macintosh, B., & Nadeau, D. 2006, ApJ, 641, 556
Marois, C., Lafrenière, D., Macintosh, B., & Doyon, R. 2008, ApJ, 673, 647
Masciadri, E., Mundt, R., Henning, Th., Alvarez, C., & Barrado y Navascués, D. 2005, ApJ, 625, 1004
Park, S. K. & Schowengerdt, R. A. 1983, Comput. Vis. Graphics Image Process., 23, 258
Racine, R., Walker, G., Nadeau, D., Doyon, R., & Marois, C. 1999, PASP, 111, 587
Ribak, E. N. & Gladysz, S. 2008, Opt. Exp., 16, 15553
Roberts, L. C. J. et al. 2005, AJ, 130, 2262
Scott, D. W. 1979, Biometrika, 66, 605
Soummer, R. & Ferrari, A. 2007, ApJ, 663, L49
Soummer, R., Ferrari, A., Aime, C., & Jolissaint, L. 2007, ApJ, 669, 642
Sparks, W. B. & Ford, H. C. 2002, ApJ, 578, 543
Thatte, N., Abuter, R., Tecza, M., Nielsen, E. L., Clarke, F. J., & Close, L. M. 2007, MNRAS, 378, 1229